\documentclass[twocolumn,prb,superscriptaddress]{revtex4}
\usepackage{graphicx}
\usepackage{dcolumn}
\usepackage{bm}
\newcommand \ga{\raisebox{-.5ex}{$\stackrel{>}{\sim}$}}
\newcommand{\up}{ \uparrow }
\newcommand{\down}{ \downarrow }
\newcommand{\beq}{\begin{equation}}
\newcommand{\eeq}{\end{equation}}
\newcommand{\bea}{\begin{eqnarray}}
\newcommand{\eea}{\end{eqnarray}}
\begin{document}
\title{Many-body wavefunctions for normal liquid $^3$He}
\author{Markus Holzmann}
\author{Bernard Bernu}
\affiliation{LPTMC, UMR 7600 of CNRS, Universit\'e  Pierre et Marie Curie, Paris,
France}
\author{D. M. Ceperley}
\affiliation{Dept. of Physics and NCSA, U. of Illinois at
Urbana-Champaign, Urbana, IL 61801, USA}
\begin{abstract}
We present new trial wave-functions which include 3-body
correlations into the backflow coordinates and a 4-body symmetric
potential. We show that our wavefunctions lower the energy enough
to stabilize the ground state energies of normal liquid $^3$He in
the unpolarized state at all pressures in agreement with
experiment; however, quantitative discrepancies remain. Further,
we include strong spin coupling into the Fermi liquid by adapting
pairing wave functions.
We demonstrate a new, numerically stable
method  to evaluate
pairing functions
which is
also useful for Path Integrals calculations at low, but non-zero
temperatures.

\end{abstract}
\pacs{PACS: }
\maketitle

For the understanding and development of quantum many-body
systems, liquid helium is a key test case due to its
experimentally well-characterized low temperature phases and the
simplicity of the helium interaction. The calculation of its
thermodynamic and ground state properties is one of the most
important benchmarks for microscopic theory and for the quantum
Monte Carlo(QMC) simulation methods
\cite{McMillan65,Levesque65,Kalos74,Ceperley77}. While QMC can
calculate exact properties for bosonic systems, and has
successfully been applied to calculate properties for liquid and
solid $^4$He \cite{Whitlock79,rmpi},  fermion systems are still
challenging, since antisymmetry leads to the so-called fermion
sign problem. In general, no exact method is known which provides
a solution for an extended fermion quantum system in two or three
spatial dimensions.

To overcome the fermion-sign problem, most fermion QMC
calculations rely on the fixed-node (FN) approximation where the
nodes of a trial wavefunction, $\psi_T$, are imposed as a boundary
condition on the many-body Schr{\"o}dinger equation with the
projector Diffusion Monte Carlo (DMC) method \cite{Reynolds82}.
Since the nodal surfaces of the exact ground state wavefunction
are, in general unknown, the energies of FN-DMC calculations are
higher than the exact answer by a small but unknown amount.
Progress in fermionic Monte Carlo calculations is, therefore,
often connected to progress in constructing new types of trial
wavefunctions. Actually, for homogeneous quantum systems, only a
few types of different trial wavefunctions have been successfully
applied within QMC, in particular for liquid $^3$He
\cite{Lee81,Schmidt81,Panoff89,Casulleras00}.

Many calculations use the simplest possible nodal structure based
on a Slater-Jastrow wavefunction \beq \psi_{SJ}=
\det_{ki}\phi_k({\bf r}_i) \exp[-U_J] \eeq where ${\bf r}_i$ are
the positions of the $N$ fermions ($i=1,\dots N$). Antisymmetry is
assured by a Slater determinant, $\det_{ki}\phi_k({\bf r}_i)$ of
single particle orbitals $\phi_k({\bf r})$, where $k$ labels one
of the $N$ single particle orbitals and the Jastrow potential
$U_J[R]=\sum_{ij}u(r_{ij})$, $r_{ij}=|{\bf r}_i-{\bf r}_j|$, takes
into account correlations ($R$ indicates its dependance on all the
particle coordinates). Since it is symmetric with respect to
particle exchange and real, it does not modify the nodes of the
many-body wavefunction and, therefore, it does not influence the
ground state energy within a FN-DMC calculation.

Homogenous quantum systems play an important role in understanding
many-body correlations, since for inhomogeneous systems, the
Hartree-energy typically dominates the correlation energy.
Translational invariance
restricts the single particle
orbitals to be plane waves with energies less than the Fermi
level. This explains why the simple Fermi liquid Slater
determinant gives accurate results, and, at the same time, it
explains the difficulty in going beyond Slater-Jastrow without
violating homogeneity.

Backflow \cite{Feynman56} is one possibility to
extend this type of wavefunction by
including many-body correlations in the nodes
\cite{Panda73,Schmidt79,Manousakis83} using a Slater
determinant $\det_{ki}\phi_k({\bf q}_i^{BF})$ where the bare
coordinates ${\bf r}_i$ used in the Slater determinant of the
Slater-Jastrow wavefunctions are replaced by dressed
``quasi-particle'' coordinates
\beq
 {\bf q}_{i}^{BF}={\bf r}_{i}+{\bf  Y}_i
\eeq
where ${\bf Y}_k[R]=\nabla_k \sum_{ij} y(r_{ij})$.
An additional symmetric 3-body potential, $U_{3B}$, is
also added to the Jastrow part of the trial function
 \beq
 U_{3B}= \lambda_T \sum_{i} {\bf W}_i{\bf W}_i
 \label{3body}
 \eeq
with ${\bf W}_k[R]={\boldmath \nabla}_k \sum_{ij} w(r_{ij})$
\cite{footnote-1}. The resulting backflow-3-body (BF-3)
wavefunction, \beq \psi_{BF3}= \det_{ki}\phi_k({\bf q}_i^{BF})
\exp\left\{ -U_J-U_{3B}\right\}, \eeq contains the unspecified
functions $u(r)$, $w(r)$, and $y(r)$; these are determined either
by analytical theories or by numerically minimizing the
expectation value of the energy using simple analytical forms with
free parameters. The backflow 3-body wavefunction (BF-3) has
provided the most accurate results for the electron-gas
\cite{Kwon93,Kwon98,BF} and for liquid $^3$He
\cite{Panoff89,Casulleras00,Moroni95,Zong03}. Spin-dependent
correlations have been considered in Ref.~\cite{Vitiello97}, but
have found to be roughly equivalent to backflow.

However, the calculated liquid $^3$He energy and magnetic
susceptibility is still in disagreement with experimental
estimates \cite{Moroni95,Zong03}; the FN-DMC ground state energy
is higher by about $260$mK. The prediction of the polarization
energy is particularly unsatisfactory: within Variational Monte
Carlo (VMC), the Slater-Jastrow wavefunction predicts a polarized
fluid at zero pressure \cite{Levesque80}, inclusion of three-body
correlations results in a slight improvement, however, still
largely unsatisfactory \cite{Lhuillier81}. Backflow mainly affects
the unpolarized state and therefore improves the estimate of the
polarization energy. Only within FN-DMC is the unpolarized state
stable at zero pressure, but the computed susceptibility is still
too large, and at higher pressure, the unpolarized state becomes
unstable \cite{Zong03}. Because liquid $^3$He is such an important
benchmark system, it is crucial to find what correlations are
absent in the backflow-3-body wavefunctions.

Pairing wavefunctions, where the Slater determinant is replaced by
the antisymmetrization of a pairing wavefunction $\phi(r_i,r_j)$
has been suggested by Bouchaud and Lhuillier \cite{Bouchaud88} to
overcome this problem. In general, the antisymmetrization of a
pairing functions leads to a pfaffian,
$\mathrm{Pf}_{i,j}\phi(r_i,r_j)$ which reduces to a determinant
for a $M=0$ (s-wave) pairing wavefunction.
Recently, singlet 
pairing wavefunctions, have been applied to calculations on the
BEC-BCS crossover of fermionic gases
\cite{Carlson03,Chang04,Giorgini}. As far as the calculations on
$^3$He are concerned, spin-triplet pairing wavefunction with $M=1$
(p-wave) are promising, but it remains to be shown that they
indeed provide lower energies than the backflow-3 body
wavefunctions.

It is known experimentally that the ground state of $^3$He is a
{\em superfluid} with {\em spin triplet pairing}. The transition
occurs around $1$mK which is an estimate of the energy gap,
$\Delta$. We expect that energy differences between the true
superfluid ground state and the best Fermi liquid state are small,
of order $\Delta \sim 1$mK and the introduction of a superfluid
pairing wavefunction alone {\em can not resolve} the {\em
energetic mismatch} between theory and experiment of order
$260$mK. Further, one expects that pairing only involves states
close to the Fermi surface, e.g. states of momenta $k$ with
$k_F-\delta k<k< k_F+\delta k$ where $k=k_F$ at the Fermi surface
and $\delta k \sim 2 m \Delta /k_F$ defines the typical coherence
length $\xi \sim \delta k^{-1}$ of the pairing wave function. In
$^3$He ($k_F\sim 0.9\AA^{-1}$)  one estimates $\delta k \sim
10^{-2}\AA^{-1}$, and the coherence length $\xi \sim 100 \AA$
exceeds by an order of magnitude the size of a typical simulation
box, $L$, as $L\sim 10 \AA$ for a system of $N=66$ atoms. Strong
correlations in helium in changing the bare mass into the
effective mass hardly modify this conclusion. On the other hand
size effects (for a non-degenerate ground state) will favor the
normal state since the energy of a weakly bound state is increased
when the size of the system is decreased. For these reasons, we
mainly limit the discussion to the normal state.

\begin{table}
 \begin{tabular}{|c|c|l|l|l|} \hline
 $\rho$ (nm$^{-3}$)  & wavefunction & $E_v (K)$ & $\sigma^2$ & $E_{DMC} (K)$ \\\hline
16.35   &   BF-3     &   -2.201(6)&   23        & -2.417(1)   \\
   &   3BF-4& -2.284(3)  &14  & -2.438(1)  \\
    &  S3BF-4& -2.294(3) &14 & -2.432(4) \\
 \hline
 19.46   &   BF-3     &   -1.775(5)&   26       & -2.155(5)   \\
   &   3BF-4& -1.905(4)  &20  & -2.174(3)  \\
    &  S3BF-4& -1.945(4) &20 & -2.182(4)  \\
 \hline
  23.80  &   BF-3    &     -0.055(1)      & 65   & -0.77 (2)   \\
    &   3BF-4 & -0.272(7)  & 53& -0.834(2) \\
    &   S3BF-4 & -0.340(5) &49 & -0.861(4)\\
   \hline
\end{tabular}
\caption{Variational ($E_v$) and FN-DMC energies ($E_{DMC}$) in
K/atom for a system of $N=66$ unpolarized $^3$He atoms with
periodic boundary conditions at two different densities;
$\sigma^2$ is the variance per atom; BF-3 are
results of the backflow-3body  
wavefunction, 3BF-4 are the results
presented here using an additional 3-body correlation for the
backflow and an additional symmetric 4-body term, and S3BF-4 is
the pairing form described below.}
\end{table}

In this paper, we propose new types of trial wavefunctions with
explicit many-body character and use them to calculate the
unpolarized ground state energy of $^3$He at the maximum,
intermediate, and the minimum densities of liquid $^3$He. First,
we introduce  a three-particle backflow and 4-body correlations
(3BF-4), as a natural extension of the backflow-3-body (BF-3)
Fermi liquid wavefunction.  As in the case of backflow, the
Slater-determinant is a function of quasi-particle coordinates,
but a three-body correlation is added to the original backflow
coordinate \beq
    {\bf q}_{i}^{3BF}={\bf q}_{i}^{BF}+\hat{{\bf Q}}_i{\bf D}_i
\eeq
where $\hat{{\bf Q}}_i [R]$ is a tensor with generic elements
 $\hat{{\bf Q}}_k [R])_{\alpha \beta} =\nabla_{k\alpha}
\nabla_{k\beta} \sum_{ij} q(r_{ij}) +\delta_{\alpha \beta}
\nabla_k^2 \sum_{ij} q'(r_{ij})$, and ${\bf D}_k[R] = \nabla_k
\sum_{ij} d(r_{ij})$ is a vector. Furthermore, a symmetric 4-body
potential  is added to the Jastrow part of the trial function:
\beq U_{4B} = \sum_{i} {\bf X}_i \hat{\bf P}_i {\bf X}_i
\label{4body} \eeq where $\hat{\bf P}_i[R]$ is a tensor and ${\bf
X}_i $ a vector, analogous to $\hat{\bf Q}$ and $\bf D$. The new trial
wavefunction including 3-body backflow and a 4-body potential
(3BF-4) is:
 \beq
 \psi_{3BF4}=\det_{ki} \phi_{k}({\bf q}_i^{3BF})
 \exp[-U_J-U_{3B}-U_{4B}].
 \eeq

As in previous studies, we use gaussian functions as a basis for
the 1-D functions arising in the wavefunction (such as $q$,$q'$
and $d$), and minimize a combination of energy and variance of the
trial wavefunction within VMC. Results for a system of $N=66$
unpolarized $^3$He atoms interacting with the recent Aziz-SAPT2
potential \cite{Korona97,Janzen97} with periodic boundary
conditions are shown in table I \cite{footnote-2}.
To reduce finite size effects we have also performed twist
averaging  \cite{Lin01} and we have corrected size-effects in the
potential energy. Results for different polarizations are given in
table II. 
For comparision we have added the twist-averaged results of the backflow
3-body wavefunctions of Ref.~\cite{Zong03} in table II and the
corresponding results for periodic boundary conditions in table I.

The three densities $16.35$, $19.46$, and $23.80$ atoms
$nm^{-3}$ correspond to pressures equal to $0$, $8.4$, and $35.3$
bar. The magnetic susceptibility, $\chi/C$, has been estimated 
in table III by
assuming $E=E_0 +\zeta^2/(2\chi/C) + a_4 \zeta^4$ where
$\zeta$ is the spin polarization. At the
equilibrium density, inclusion of 3 and 4-body correlations reduce
the magnetic susceptibility in agreement with the experimental
value \cite{Goudon}. Away from equilibrium density, $a_4$ becomes
more important, and the estimate of $\chi/C$ at the intermediate
density is less reliable. Close to crystallization, the energy
variation between the unpolarized and the partially polarized
state for the 3BF-4 wavefunction is of the order of the
uncertainty of the calculation.
We have not performed calculations for the 100\% polarized phase
since backflow is already much less important than in the
unpolarized phase \cite{Vitiello97}; similarly 3-body correlations
mainly affect the unpolarized state \cite{Lhuillier81}. Therefore,
we do not expect significant corrections to the energies of the
complete polarized state of the BF-3 trial function.

We have tried also different tensorial forms  for a 4-body trial
function, a 5-body trial function, as well as 4-body and
k-dependent corrections to the quasi-particle coordinates without
significant improvements ($<5$mK). We cannot exclude the
possibility that more effective optimization with more flexible
functions would lead to significantly lower energy. This possibility,
together with the application of the released node technique
\cite{release} is
under current investigation.

\begin{table}
\begin{tabular}{|c|c|cc|cc|} \hline
 $\rho$   &$\zeta$&$E_v(K)$ &$E_{DMC}(K)$ & $E_v(K)$ &$E_{DMC}(K)$\\
(nm$^{-3}$) & & BF-3 & BF-3& 3BF-4& 3BF-4\\\hline
  16.35  &0&  -2.1633(9)&-2.3586(8)&   -2.241(1)&   -2.3802(4)  \\
      &0.242&  -2.1614(9)&-2.3548(10)&    -2.242(1)&  -2.3743(1)    \\
         &0.485&  -2.1753(9)&-2.3433(8)&  -2.248(1)&  -2.3633(6)  \\\hline
  19.46 &0& -1.7469(12)& -2.0685(13)&    -1.863(1)&   -2.0972(9)  \\
    &0.242& -1.7499(12)&-2.0623(13)&    -1.873(1)&   -2.0948(6)  \\
 &0.485& -1.7984(12)&-2.0807(11)&    -1.905(1)&   -2.1046(8)  \\\hline
   23.80     &0&  -0.0184(20)& -0.6612(20)& -0.203(2) &  -0.7094(9) \\
   &0.242& -0.0229(21)& -0.6619(21)& -0.222(2) &  -0.7093(11)\\\hline
\end{tabular}
 \caption{VMC and DMC energies  of the BF-3 wavefunction of 
 Ref.~.\cite{Zong03} and of the
 3BF-4 wavefunction using twist averaged boundary
 conditions. Notations and units as in table I;
 $\zeta$ is the polarization.}
\end{table}
\begin{table}
\begin{tabular}{|c|c|c||c|c|} \hline
 $\rho$  & $\chi/C (K^{-1})$& $\chi/C (K^{-1})$ & $\rho_{\mathrm exp}$    & $\chi_{\mathrm exp}/C$\\
 (nm$^{-3}$) &  BF-3 & 3BF-4 &  (nm$^{-3}$) & ($K^{-1}$) \\\hline
  16.35  & 8(2) & 4.5(5)&16.37 & 3.0 \\\hline
  19.46 & 5(2) & 8(3) & 19.44 & 4.0 \\\hline
   23.80& -   & $|\chi/C| \ga14$& 23.45 & 6.1\\\hline
\end{tabular}
 \caption{ Magnetic susceptibility 
 $\chi/C$  in $K^{-1}$, where $C$ is the molar Curie constant, estimated from
 the DMC calculations using the 3BF-4 wavefunction;  
  experimental values \cite{Goudon} at densities $\rho_{\mathrm exp}$ are given
  for comparision.}
\end{table}

We have also considered pairing wavefunctions. Simple functional
forms (Gaussians) for the pairing functions (plus  the Jastrow
potential) have led to energies considerably higher than those from the
BF-3 trial functions. Shell effects lead to large size effects and
make a direct comparison of energies of triplet and singlet
pairing wavefunctions problematic without a careful extrapolation
to the thermodynamic limit. Since triplet pairing functions did
not show significant improvements over the singlet pairing, we
restrict the pairing functions to spin-singlet pairing: the
pfaffian reduces to a determinant and the antisymmetric part of
the wavefunction is:
 \beq
 \det_{ij} \phi_s({\bf r}_{i\up}-{\bf r}_{j\down})
 \label{singlet}
 \eeq
where ${\bf r}_{i\up}$ (${\bf r}_{i\down}$) represents the
coordinates of particle $i$ with spin-up (down). Using a Gaussian
form for $\phi_s(r)$, we have that the minimum energy orbital has
a width as large as the simulation cell, $L$. In this case, the
orbitals have to be periodized, which is most easily done using a
Fourier sum. Optimization of the Fourier coefficients led to the
limiting form of the Fermi liquid:
 \beq \phi_{FL}(r)=\sum_{k \le k_F} e^{i{\bf k} \cdot {\bf r}}
 \label{SFermi}
 \eeq
We conclude that pairing of bounded singlet/triplet pairs does not
improve the energy significantly, consistent with the small
pairing energy in superfluid $^3$He. \cite{footnote-3}

To go beyond the Fermi liquid trial function, we can use the
singlet pairing wavefunction as a way of introducing strong
spin-correlations:
 \beq
 \phi_s(r)=\sum_{k \le k_F} e^{i {\bf k} \cdot {\bf  r} [1 +\eta(r)]}
 e^{-\nu(r)}
 \label{Sused}
 \eeq
in the determinant of Eq.~(\ref{singlet}). Here $\eta(r)$ and
$\nu(r)$ are localized functions vanishing at large particle
separation. They introduce modulations  and correlations in the
pairing of opposite-spin atoms without affecting the long-range
correlations of the Fermi-liquid determinant. They do not describe
a bound state, so we do not expect them to describe a superfluid
state, but they might be a precursor to superfluidity. However,
the correlations between unlike spins turn out to be much stronger
than introducing a spin-dependent backflow or Jastrow potential
which have not been successful in lowering the energy. We also
include backflow, 3-body backflow, 3-body, and 4-body potentials
as discussed above: \beq \psi_{S3BF4}=\det_{i,j} \phi_s({\bf q}_{i
\up}^{3BF} \!\! -\!\!{\bf q}_{j\down}^{3BF})
e^{-U_J-U_{3B}-U_{4B}}. \label{S3BF4} \eeq As shown in table I,
the energies are lowered by this wavefunction. By construction,
this wavefunction does not affect the energy of the completely
polarized state and, therefore, it stabilizes the unpolarized
phase. We have not performed twist-averaging,
but it is likely that the relative gain of the energy compared to
the 3-body backflow wavefunction survives in the thermodynamic
limit, but  a more detailed study of the finite size extrapolation
of this wavefunction remains to be done.
In Ref. \cite{Zong03}, it was estimated that the 3-body potential 
together with the effect of higher order many body terms in the interparticle
interactions raise the energy by $\sim 140$mK, and have to be added
to the computed results when compared to the experimental energy\cite{energy}
$-2.481 K$ at $\rho=16.35$nm$^{-3}$.
 Even if the absolute
energies are therefore still in disagreement with experiments by $\sim
240$mK
the  agreement with
the compressibility is considerably improved. 
Further, we expect
a reduced magnetic susceptibility compared to the
3BF-4 wavefunctions, closer to the experimental values. At high density
$10-20$mK might be still missing to  stabilize completely the unpolarized
phase. However, size effects are of the order of $30$mK, and a
clear prediction would involve an extensive finite size study.

We briefly explain our method to evaluate the determinant of the
pairing function in Eq.({\ref{S3BF4}).  A straightforward
evaluation of the determinant of the matrix
$\Phi_{ij}=\phi_s(r_i,r_j)$ is numerically unstable in the limit
where the wavefunction approaches a Fermi liquid,
Eq.~(\ref{SFermi}), because it  becomes ill-conditioned; a direct evaluation
of the determinant is dominated by round-off errors as the number
of particles, $N$, increases. Let us consider the expansion of the
pairing functions, into eigen-modes, $\varphi_k(r)$, \beq
\phi_s(r_i,r_j)=\sum_k d_k \varphi^*_k(r_i) \varphi_k(r_j)
\label{PAIR} \eeq where $d_k$ is the occupation number of
eigenstate $k$. For a Fermi liquid, the summation is dominated by
$N$ modes, $k=1,\dots,N$. We introduce the matrices of the
Slater-determinants of the Fermi Liquid,
$L_{ik}=\varphi^*_k(r_i)$, $R_{ki}=\varphi_k(r_i)$, and a diagonal
matrix $D_{kk'}=\delta_{k,k'} d_k$, and write Eq.(\ref{PAIR}) as a
matrix, \beq \Phi=LDR+M =L D^{1/2} \left( 1 + \tilde M  \right)
D^{1/2}R \label{NEWALG} \eeq with $M_{ij}\!=\! \sum_{k>N}\! d_k
\varphi^*_k(r_i) \varphi_k(r_j)$, and $\tilde{M}\!=\!L^{-1}
D^{-1/2}M D^{-1/2} R^{-1}$. Now, the determinant of
Eq.(\ref{NEWALG}) can be written as: \beq \det \Phi  =   \det L
\det D \det R
  \det (  1 + \tilde M )
  \label{detphi}
\eeq
and involves only the usual Slater determinants, $\det L$,
$\det R$ and their inverses which are well-behaved. The matrix
$\tilde M$ is typically much more rapidly decaying in space and a
numerical evaluation of the terms on the rhs of Eq.~(\ref{detphi})
is numerically stable,  including all limiting cases.  This method is not
restricted to spin singlet/triplet pairing, but it is also useful
to calculate finite temperature density matrices well below the
Fermi-temperature where the same problem of extremely ill
conditioned determinants occurs.

In conclusion, we have found 3-body and 4-body correlations
improving the energy and the magnetic susceptibility of liquid
$^3$He. We expect that the new terms will be important to
calculate the Fermi liquid parameters in strongly correlated
liquids such as liquid $^3$He and the electron gas, the
polarization transition of the electron gas close to Wigner
crystallization and related systems. The method to calculate
pairing function close to the Fermi liquid should further allow
the study  the very dilute limit of a BCS gas.

{\em Acknowledgments}.
 Computer time has been provided by CNRS-IDRIS. Support for DMC was
 NSF DMR 04-04853. MH acknowledges discussions with C. Lhuillier,
 V. Goudon, and H. Godfrin, and hospitality of LP2MC (Grenoble).


\begin{thebibliography}{10}
\bibitem{footnote-1} The 3-body
potential of Eq.~(\ref{3body}) includes ``self-terms'' of the form
$\sum_{ij} [ \nabla_i w(r_{ij}) ]^2$ which should be subtracted
out. To simplify the notation self terms are assumed to be
subtracted in the following.
\bibitem{footnote-3}
Finite size effects
are important as one has to express the pairing function in
Fourier space for weakly bound pairs and the discrete level
spacing of the finite system  dominates.
\end{thebibliography}
\end{document}